\documentclass[preprint2]{aastex}
\begin{document}

\title{The Polytropic Equation of State of Primordial Gas Clouds}

\author{Marco Spaans\altaffilmark{1} and Joseph Silk\altaffilmark{2}}

\affil{$^1$Kapteyn Astronomical Institute, P.O. Box 800, 9700 AV
Groningen, The Netherlands; spaans@astro.rug.nl}
\affil{$^2$Astrophysics, Oxford University, 1 Keble Road, Oxford OX1
3RH, United Kingdom; silk@astro.ox.ac.uk}

\begin{abstract}

The polytropic equation of state (EOS) of primordial gas clouds with
modest enrichment is computed, motivated by the recent observations of
very Fe-deficient stars, [Fe/H]$\sim 10^{-3.5}-10^{-5}$, such as  HE
0107-5240 and CS 29498-043. These stars are overabundant, relative to
Fe, in C and O.  We assume that the observed abundances of species
like C, O, Si and Fe are representative of the gas from which the
currently observed metal-deficient stars formed.  Under this
assumption, we find that this primordial metal abundance pattern has
profound consequences for the thermal balance and chemical composition
of the gas, and hence for the EOS of the parental cloud.
The polytopic EOS is soft for low, [O/H]$<10^{-3}$, oxygen abundances,
but stiffens to a polytropic index $\gamma$ large than unity for
[O/H]$>10^{-2}$ due to the large opacity in the CO and H$_2$O cooling
lines.  It is further found that a regulating role is played by the
presence and temperature of the dust, even when the overall carbon
abundance is only [C/H]$\sim 10^{-2}$. When the dust is warmer than
the gas, a region with $\gamma\sim 1.2$ results around a density of
$\sim 10^4$ cm$^{-3}$. When the dust is colder than the gas, a region
with $\gamma\sim 0.8$ is found for a density of $\sim 10^6$ cm$^{-3}$.
Implications for the primordial initial mass function (IMF) as well as
the IMF in starburst galaxies, where the metallicity is super-solar,
are explored and related to processes that influence the temperature
of the ambient dust.

\end{abstract}


\section{Introduction}

 A fundamental issue in the study of star formation is  
understanding  the physical
structure of molecular gas clouds from which stars are formed.
Despite a wealth of observational data for the Milky Way (c.f.\ Fuller \&
Myers 1992; Goodman et al.\ 1998), the nature of the equation of state (EOS)
 remains a major
theoretical problem in understanding the stabillity and collapse of molecular
clouds. The stiffness of the EOS can be largely responsible for the resulting
density probability function of interstellar gas in the turbulent ISM.
In particular, the value of the polytropic index $\gamma$ strongly influences
the mass distribution of density condensations as well as the amount of
clump fragmentation (V\'azquez-Semadeni et al.\ 1996; Li et al.\ 2003).
In Spaans \& Silk (2000) the properties of a polytropic EOS were investigated
and it was found that the stiffness of the EOS depends strongly on the ambient
metallicity.

Recent observations of metal-deficient stars such as HE
0107-5240 (Christlieb et al.\ 2004) and CS 29498-043 (Aoki et al.\
2004) indicate that formation of (low-mass) stars is possible in
very low metallity gas clouds.  The formation of stars in
low-metallicity environments (Abel, Bryan \& Norman, 2002) is
important for the reionization of the universe and the early
enrichment of the interstellar and intergalactic medium (Hirashita \&
Ferrara 2002). Even though more (but not many more) stars are known
with very low Fe abundances (e.g., CS 22949-037; Aoki et al.\ 2004),
we concentrate on the two stars mentioned above since they appear to
reflect the wide range in C, N, O and $\alpha$ element abundance
patterns for Fe-deficient stars.

It is not generally known what influence the environment has on the
shape of the IMF. HST/WFPC2 observations of the LMC by Gouliermis,
Brander \& Henning (2004) indicate that local conditions seem to favor
the formation of higher mass stars (top-heavy IMF) in associations,
and not in the background field.  However, Meyer et al.\ (2004, and
references therein) find that the IMF is not a strong function of
environment or initial conditions.  In any case, the numerical
simulations of Li et al.\ (2003, and references therein) indicate
that the value of the polytropic index strongly influences the
spectrum of density condensations that is formed under the influence
of gravity and turbulent driving. In particular, fragmentation is
enhanced or inhibited when $\gamma$ is smaller (0.7) or larger (1.1)
than unity. In this context, we present computations of the polytropic
EOS of primordial (low-metallicity) gas clouds at high redshifts. We
also comment on interstellar regions that have super-solar
metallicities and are  exposed to a warm (dusty) infrared
background (e.g., starbursts).

\section{Model Description}

We assume that the observed abundances of species like C, O, Si and Fe are
representative of the gas from which the currently observed metal-deficient
stars formed. We further employ a polytropic EOS,
$P\propto\rho^{\gamma}$, where $\gamma$ is the
polytropic index and $\rho$ the mass density.
We adopt a perfect gas equation of state, $P\propto\rho T_g$ for the gas
temperature $T_g$, to write $\gamma$ as
$$\gamma =1+{{d{\rm log}T_g}\over{d{\rm log}\rho}}.$$
This last step is justified (Scalo \& Biswas 2002; V\'azquez-Semadeni, Passot
\& Pouquet, 1996) as long as the heating and cooling terms in the
fluid energy equation can adjust to balance each other on a time-scale shorter
than the time-scale of the gas dynamics (i.e., thermal equilibrium).
It should be noted that because $\gamma$
depends on the (logarithmic) derivative of the temperature with
respect to density, it implicitly depends on radiative transfer effects and
changes in chemical composition through derivatives of the heating and
cooling functions (Spaans \& Silk 2000).
The model described in Spaans \& Silk (2000) is used in this work and
the interested reader is referred to that paper
for a detailed description of the various heating and cooling terms that
influence the polytropic index. The (updated) two-dimensional numerical
radiative transfer code of Spaans (1996) with the detailed chemical treatment
of Spaans \& van Dishoeck (1997), and its extensions to lower metallicities
by Norman \& Spaans (1997) and Spaans \& Norman (1997) is used in this work.
With the work of Scalo (1998) and Le Bourlot et al.\ (1999) it is possible
to extend the Spaans \& Silk (2000) model to metallicities much smaller than
1\% of solar, where H$_2$ and HD completely dominate the cooling of the gas.
The main features of the model are summarized below.

We consider self-gravitating {\it spherical} clouds and therefore
we adopt a singular isothermal sphere (Neufeld, Lepp \& Melnick 1995),
for which the total hydrogen number density
$n_{\rm H}$ scales with column density $N_{\rm H}$ per unit of velocity as
$$N_{\rm H}=7.2\times 10^{19}n_{\rm H}^{0.5} {\rm cm}^{-2}/({\rm km s}^{-1}).$$
We include number
densities, $n_{\rm H}=n(H)+2n({\rm H}_2)$, upto $2\times 10^7$ cm$^{-3}$ and
metallicities as low as $3\times 10^{-4}$ and 0 in solar units.
Furthermore, dust temperatures between 10 and 100 K can be treated and the
latest chemical reaction rates
have been used in the computations (Le Teuff, Millar \& Markwick, 2000).
Care has been taken to check the convergence of the chemical equilibrium
calculations for O/C ratios that significantly differ from solar.
In general, the thermal and chemical balance was solved  iteratively with
a convergence criterion of 0.1\% for the chemical abundances and level
populations.

The computed grid is accurate enough to do linear interpolations
between the log of any two adjacent values.
Extrapolation of the values of $\gamma$ to densities lower than
$10^2$ cm$^{-3}$ (say to $1$ cm$^{-3}$) is reliable because the slopes are
modest. Note that softening of the EOS due to gas cooling on
cold dust can be important and flips over to gas heating when the dust is warm.

We have used a cosmic ray heating rate for the low-metallicity
models that is given by a cosmic ray ionization rate of ${{1}\over{3}}\zeta$,
with $\zeta=3\times 10^{-17}$ s$^{-1}$. This means that cosmic ray
heating is provided by a background star formation rate of the order of
$1 M_\odot/{\rm yr}$.
However, the zero metallicity polytropic index is only weakly dependent on
temperature between
100 and 2000 K so that the actual value of the cosmic ray heating rate, as
long as it is proportional to density, does not matter much for the
low-metallicity case (Scalo \& Biswas 2002). For starburst environments,
we assume a cosmic ray ionization rate of $30\zeta$.

Solar relative abundances are assumed (Asplund et al., 2004; Jenkins 2004).
The latter condition
is replaced by the observed abundance pattern of metal-deficient stars for
some of the primordial gas cloud models. We further adopt a MRN grain size
distribution (Mathis, Rumpl \& Nordsieck 1977) and
assume that the dust abundance scales with the carbon abundance.

The velocity dispersion of microscopic turbulence is
$\Delta V_{\rm tur}=0.5$ km/s in the quiescent primordial gas, to which the
purely thermal contribution is
added (in the square) while the iterations on the thermal and chemical balance
are performed. However,
for the starburst model $\Delta V_{\rm tur}=3$ km/s is adopted
to take the larger input of kinetic energy (e.g., through supernovae) into
account.

No freeze-out of molecules is assumed. The latter assumption
is justified because our dust is generally warmer than 20 K. The total
extinction through the starburst region containing the model cloud is 100 mag
(Spaans \& Silk 2000).
The results of Le Bourlot et al.\ (1999) are
used for the collisional excitation of H$_2$ by H, He and H$_2$, where non-local
thermodynamic equilibrium level populations and quantum-mechanical cross
sections are computed. For HD cooling, the results of Flower et al.\ (1999)
are used.
Both H$_2$ and HD line emission are optically thin.
For completeness we have also included H$_2$ formation heating with equal
contributions from the binding energy (4.93 eV) to translational, rotational
and (statistically distributed) vibrational degrees of freedom (Sternberg \&
Dalgarno 1989; Meijerink \& Spaans 2004).

The computed values of $\gamma$ depend sensitively on a number of effects:
1) line trapping, 2) metallicity, 3) H$_2$O heating and cooling,
4) microturbulence versus a large velocity gradient (LVG) velocity field.
In particular,  4) is important in the context of collapsing molecular cloud
cores. Note that a LVG approximation suppresses line trapping somewhat, and
yields a relatively softer equation of state (see Spaans \& Silk 2000).
Also, H$_2$O heating, where absorption of infrared photons and collisional
de-excitation causes heating (Takahashi, Hollenbach \& Silk 1983), plays a
smaller role for a LVG model unless $A_V > 20$ mag.

We assume a microturbulent velocity field for all simulations.
For optically thin (e.g., low-metallicity) line emission the velocity field
plays no role. We ignore time-dependent effects (Bromm \& Loeb 2003).
UV illumination from the outside, not included here, will create a warm PDR
region where the cooling is dominated by [CII] and [OI] fine-structure emission
and the polytropic index varies between 0.5-1.0 for C$^+$ and
between 0.6-1.2 for O-dominated emission.

\begin{table*}
\caption{All models use $\Delta V_{\rm tur}=0.5$ km/s, except for model 14, which uses $\Delta V_{\rm tur}=3$ km/s. The observed O and C abundance for the halo stars HE 0107-5240 and CS 29498-043 are adopted in models A, B1 and B2. For CS 29498-043 the two adopted C abundances are indicative of the observational uncertainties in carbon.}
\begin{tabular}{|c|c|c|c|c|c|c|}
\hline
C [$Z_\odot$] &O/C [O$_\odot$/C$_\odot$] &z &Dust &Model &Figure &Color\\
\hline
\hline
0    &0   &-  &no  &1  &1-4 &red\\
0.03 &1/3 &10 &no  &2  &1 &light blue\\
0.03 &1/3 &20 &no  &3  &1 &blue\\
0.01 &1/3 &10 &no  &4  &1 &black\\
0.01 &1/3 &20 &no  &5  &1 &green\\
0.03 &1   &10 &yes &6  &2 &light blue\\
0.03 &1   &20 &yes &7  &2 &blue\\
0.01 &1   &10 &yes &8  &2 &black\\
0.01 &1   &20 &yes &9  &2 &green\\
0.03 &0   &10 &no  &10 &3 &light blue\\
0.03 &0   &20 &no  &11 &3 &blue\\
0.01 &0   &20 &no  &12 &3 &green\\
$3\times 10^{-4}$ &0 &0 &no &13 &3 &black\\
2    &1   &0  &yes, 75 K &14, starburst &4 &black\\
0.03 &O at 0.0013$Z_\odot$ &10 &no &A, HE 0107-5240 &4 &light blue\\
0.1  &3   &20 &yes &B1, CS 29498-043 &4 &blue\\
0.03 &3   &0  &yes &B2, CS 29498-043 &4 &green\\
\hline
\end{tabular}
\end{table*}

\section{Model Results}

In order to assess the impact of elemental abundances that strongly
deviate from solar, a grid was run with the parameters listed in Table 1.
The figure number and color coding of each model are listed.
The primordial model 1 is used as a baseline to assess the impact
of C, O and dust on the polytropic EOS.
Redshifts of 10 and 20 are considered for the high redshift universe
and the temperature of the dust, if any is present, is taken equal to that
of the CMB.
Overall, the grid covers the two Fe-deficient stars, through models A, B1 and
B2, and a range in carbon
that limits, through model 13, to the zero-metallicity primordial cooling
curve. Furthermore, various oxygen models are explored to show the large
impact that O has on the EOS, i.e., models 10-12 for no O and models
2-9 for solar and super-solar O/C ratios. The latter are relevant because
observations of CS 29498-043 (Aoki et al.\ 2004) indicate that the oxygen over
carbon ratio can also be enhanced relative to the solar ratio.
Finally, model 14 is
a starburst environment where O/C is solar, but the overall metallicity is
super-solar by a factor of two.

It is straightforward to identify a number of trends in the models.
We first note that in all models with a non-zero oxygen abundance, part of
the gas cooling is provided by the [OI] 63 um line. This
line can go into absorption against the warm CMB below a gas kinetic
temperature of $\sim 80$ K. Collisional de-excitation of excited [OI] then
leads to heating and a change in the polytropic index with the CMB temperature,
e.g., in models 2-5, even when no dust is present. The same effect occurs for
absorption of the CMB by water, i.e., H$_2$O collisional de-excitation heating.
From figures 1-4 one sees further that:
i) Carbon alone, even at 3\% of solar, is not
able to significantly stiffen ($\gamma > 1$) the EOS (figure 3).
ii) Oxygen needs to be
present at no less than the 1\% of solar level to boost the opacity
in the CO and H$_2$O lines at densities
around $3\times 10^3-3\times 10^4$ cm$^{-3}$, see figure 1, and
yield values of $\gamma$ larger than unity. Note in this that the total
abundance of oxygen (rather than carbon) sets a limit to the maximum amount of
gas phase CO when [O/H]$<1/2$[C/H].
iii) The model parental cloud of HE 0107-5240 is characterized by a soft EOS,
while that of CS 29498-043 has a stiff EOS, see figure 4. iv) An
overabundance (relative to solar) of oxygen leads to a significantly
stiffer EOS (figure 2), even when
the overall metallicity is only of the order of 3\% of solar.
That is, if the O/C ratio is enhanced by a factor of three compared to its
solar value, then the EOS stiffens to values of $\gamma$ in the 1.2 range
at around $n_{\rm H}\sim 10^4$ cm$^{-3}$. v) This effect is further
enhanced (also figure 2) when
the temperature of the dust is high, due to the CMB at $z=10-20$,
and allows H$_2$O (collisional de-excitation)
heating to dominate the thermal budget and cause the gas
heating to attain a $n_{\rm H}^2$ density dependence. vi) At densities around
$10^6$ cm$^{-3}$, the polytropic index dips below unity
if the dust temperature, $T_d$, is lower than the gas temperature.
This results in $\gamma\sim 0.8$ for models with dust at $z=0$.
If the dust grains are warm, then dust-grain heating causes the EOS
to stiffen further and $\gamma\approx 1.1$ beyond the H$_2$O opacity peak.
Recall in this case that gas-dust heating/cooling scales as
$(T_g-T_d)T_g^{0.5}n_{\rm H}^2$. vii) The starburst (figure 4) has
the stiffest EOS because of its high absolute metallicity and warm
($T_d=75$ K) dust. viii) The temperature of the dust, if present, plays an
important regulating role, see figure 2.

\begin{figure}
\epsscale{.95}
\plotone{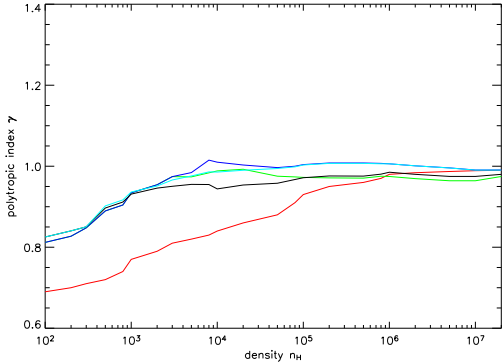}
\caption{Figure 1 shows models 1-5; note that $\gamma$ stiffens to just above unity around $3000$ cm$^{-3}$ for an oxygen abundance of $0.01Z_\odot$.\label{}}
\end{figure}

\begin{figure}
\epsscale{.95}
\plotone{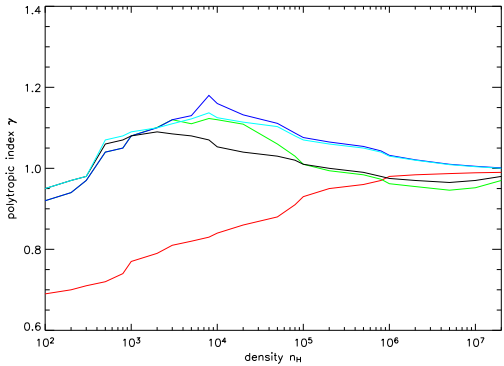}
\caption{Figure 2 shows models 1 and 6-9; note that $\gamma$ is well above unity for densities above $3000$ cm$^{-3}$, driven by the oxygen abundance.\label{}}
\end{figure}

\begin{figure}
\epsscale{.95}
\plotone{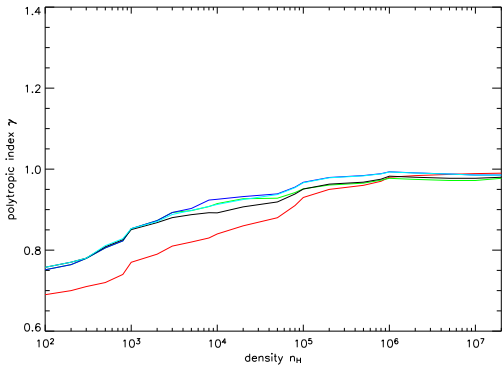}
\caption{Figure 3 shows models 1 and 10-13; note that $\gamma$ remains below unity for densities upto $\sim 10^6$ cm$^{-3}$.\label{}}
\end{figure}

\begin{figure}
\epsscale{.95}
\plotone{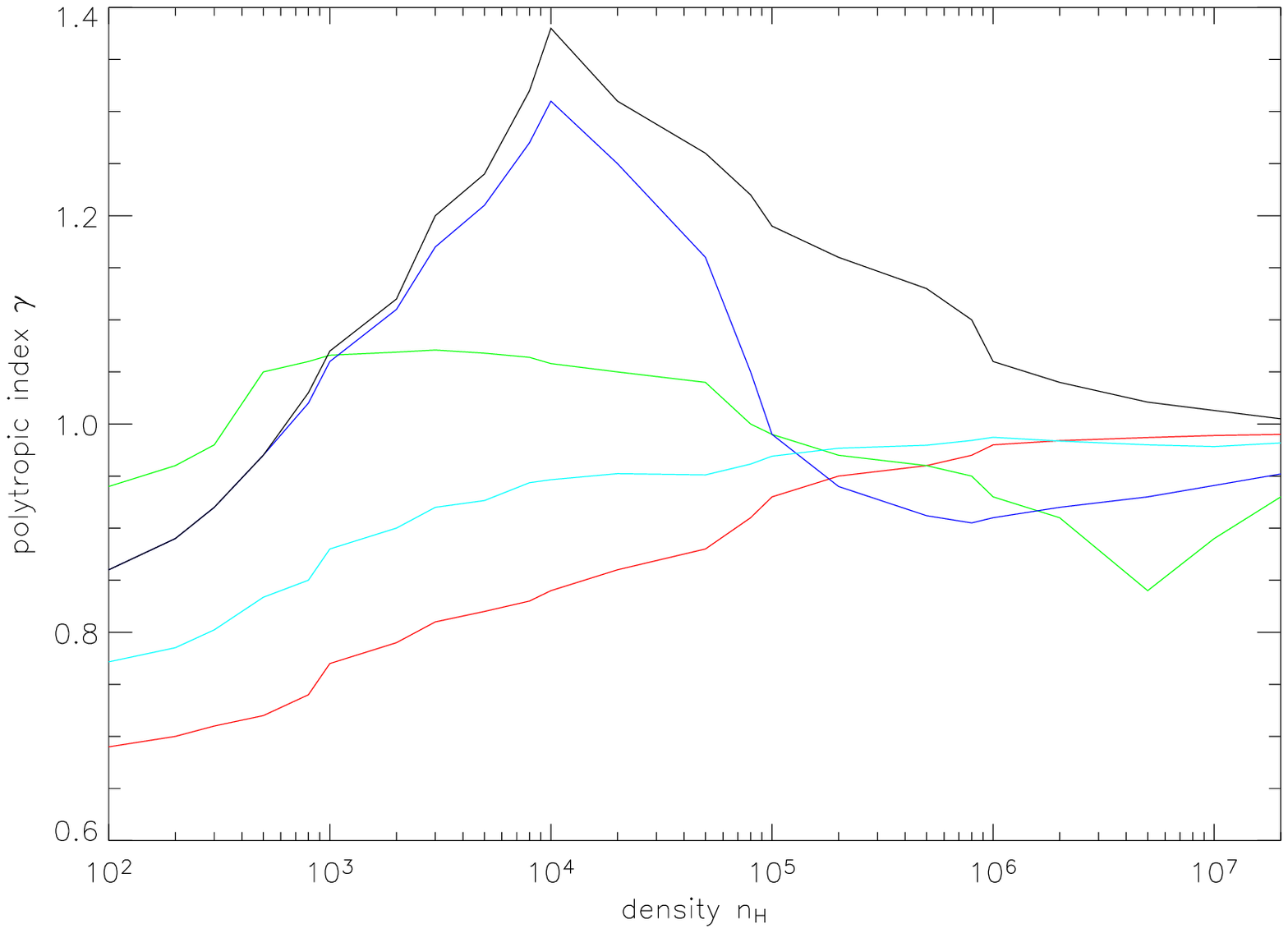}
\caption{Figure 4 shows models 1, 14, A, B1 and B2; note the progression in $\gamma$ as the oxygen abundance rises and the dust temperature is larger or smaller than the gas temperature.\label{}}
\end{figure}

\section{Discussion}

In commenting on  our results, we wish  to distinguish slightly enriched,
$Z/Z_\odot\sim 10^{-2}$, primordial gas that  leads to Fe-deficient stars
(models 2-12, A, B1 and B2) from almost zero metallicity,
$Z/Z_\odot\sim 10^{-4}$, primordial gas associated with pop III star-forming
regions (models 1 and 13). As figures 1-4 show, these two metallicity regimes
exhibit quite different behaviors as far as the polytropic index is concerned.

We point out that all the models presented in this work are
computed for a microturbulent velocity field with $\Delta V_{\rm tur}=0.5$
(or 3 for the starburst model 14) km/s and a
thermal velocity contribution that is determined by the thermal and chemical
balance of the medium. However, if large velocity gradients are present then
optical depth effects, particularly line trapping in H$_2$O and CO transitions,
are more modest and $\gamma$ decreases (Spaans \& Silk 2000).

Also, adiabatic (compressional) heating has a density dependence of
$\Gamma_{\rm A}\propto n_{\rm H}^{1.5}$, in between cosmic ray ionization,
$n_{\rm H}^{1.0}$, and gas-dust or H$_2$O, $n_{\rm H}^{2.0}$,
collisional heating.
Compressional work is not relevant to model clouds that can be described by
singular isothermal spheres; however, in a collapsing medium it
must play a role. Our heating terms bracket adiabatic heating when dust is
present and/or $Z\ge 10^{-3}Z_\odot$. For $Z/Z_\odot\le 3\times 10^{-4}$,
we expect $\gamma$ to attain a slightly larger asymptotic value of $1.0-1.3$
for $n_{\rm H}\ge 10^6$ cm$^{-3}$
(Scalo \& Biswas 2002; Schneider et al., 2002). At smaller densities the
higher resulting temperatures, $T_g\sim 200-800$ K
(Abel, Bryan \& Norman, 2000),
prevent $\gamma$ from increasing significantly under the $\Gamma_{\rm A}$ term
because excited atomic (e.g., O, Si) and molecular (e.g., CO, H$_2$O)
states with progressively higher critical densities become accessible, already
at a metallicity of $10^{-4}Z_\odot$. These higher critical densities ensure
that the sub-thermal $n_{\rm H}^2$ density dependence is maintained for larger
gas densities.

\subsection{Fe-Deficient Stars}

In Spaans \& Silk (2000), it is argued that a defining criterion for low-
or high-mass star formation to prevail, i.e., whether fragmentation can
proceed to small enough mass scales to cause a flattening in the IMF
below about 0.3 $M_\odot$ (Scalo 1998), is whether the EOS has a polytropic
index $\gamma_{\rm c}$ that
is smaller or larger than about unity, respectively. Above this canonical value
of $\gamma_{\rm c}=2(1-1/\nu )$, with $\nu $ the number of dimensions
in which cloud contraction occurs,
{\it sheets} ($\nu =2$) are able to stop fragmentation and start
gravitational contraction because the Jeans mass does not decrease any further.
In the present discussion, the thermal
Jeans mass, $M_{\rm J}\propto T^{1.5}/\rho^{0.5}$, as well as its
generalisations to turbulent and magnetic media, plays a secondary role.

If the elemental abundance
pattern that is observed for HE 0107-5240 is typical for the parental cloud in
which this star formed, then low-mass star formation is the dominating mode.
This is consistent with the modest, $0.7 M_\odot$, mass of HE 0107-5240.
Furthermore, for a solar O/C ratio, a carbon abundance of less than 3\%
of solar still results in an EOS where $\gamma$ does not become significantly
larger than unity even for warm dust grains. This would imply that the IMF
would not be top-heavy at high redshift once metal enrichment sets in. Such
an IMF would not be desirable for scenarios where the bulk of the reionization
is caused by stars. We also checked that the abundance pattern observed in
HE 0107-5240 provides sufficient levels of carbon to ensure that the cooling
time is less than the free-fall time (Bromm \& Loeb 2003).

However, if the abundance pattern of CS 29498-043 is typical, then
then we would expect that oxygen is
relatively overabundant and that the the EOS is stiff, particularly when the ambient dust
is warm. This would cause a shallow IMF at high redshift and a continuation of
the formation of high-mass stars. Such an IMF would yield a larger number of
ionizing photons per unit of mass.

In this context, it should be mentioned that theoretical models for supernova
yields of very massive, $20-130 M_\odot$,
primordial stars are capable of reproducing the C and
O abundances, but at the expense of the r- and s-process elements
(Umeda \& Nomoto 2003). An alternative would be that massive stars with
$M<8M_\odot$, i.e., stars that do not go supernova, evolve rapidly enough to
provide He-burning elements without the usual supernova dust production
(Pel, private communication). The latter scenario is interesting in that one
would have a stiff EOS around $\sim 10^4$ cm$^{-3}$, but without the ability
of cool silicate dust at $z<10$ to soften $\gamma$ around $\sim 10^6$
cm$^{-3}$. Evidence is mounting that many carbon-enriched
metal deficient stars are in binaries, where they may have picked up enriched
ejecta of an intermediate-mass companion (c.f.\ Lucatello et al., 2004). The
observed stellar abundance pattern is then still indicative of the elemental
interstellar gas composition at early times.

\subsection{Primordial Limit}

In the limit of very low metallicity, $Z/Z_\odot <3\times 10^{-4}$ or less,
the work of Schneider et al.\ (2002, 2003) and Bromm et al.\ (2001) suggests
that a metallicity of $\sim 10^{-4}Z_\odot$ at a density of $\sim 10^{12}$
cm$^{-3}$ leads to
efficient cooling and causes a softening of the EOS that gives rise to
fragmentation and the formation of clumps with masses $<1 M_\odot$.
We would like to stress that the polytropic index $\gamma$ is  already
less than
unity at much smaller, $n_{\rm H}<10^6$ cm$^{-3}$, densities 
(Scalo \& Biswas 2002) as shown by models 1 and 13.
While in the density regime around $10^{12}$ cm$^{-3}$, individual clumps are
no longer able to separate and accretion/coagulation dominates, at low
densities the higher fragment mass ($M_{\rm J}\propto\rho^{-0.5}$) may still
lead to low-mass fragmentation (e.g., through disk formation) because the
medium is very unstable to (turbulent) compression when $\gamma\approx 0.7$
(Li et al.\ 2003). In this case, a metallicity of around
$3\times 10^{-4}Z_\odot$
would still retain a soft EOS\footnote{In a medium with carbon and oxygen in atomic form, fine-structure cooling would, in fact, lower $\gamma$ to below $0.6$ at densities below $10^4$ cm$^{-3}$.},
while it would also strongly lower the ambient cloud
temperature (Bromm \& Loeb 2003) and reduce the thermal Jeans mass
($M_{\rm J}\propto T^{1.5}$) that pertains to a cloud prior to fragmentation.
Three-dimensional hydrodynamical simulations indicate that it is the
initial Jeans mass that controls, in part, the subsequent fragmentation
(V\'azquez-Semadeni et al., 1996). Following the discussion
in Spaans \& Silk (2000) on the role of the Jeans mass, we would argue that
$\gamma$, already at low densities, determines the shape of the IMF, while
the instantaneous value of the Jeans mass during fragmentation plays a lesser
role.

\subsection{Starbursts}

A similar scenario may apply to starburst galaxies. The super-solar abundances
(Barthel 2004), high gas temperatures and high densities in
those regions imply large amounts of H$_2$O.
Also, the dust (set here at 75 K) is warm due to the intense ultraviolet
radiation
field, which facilitates a) H$_2$O heating around $10^4$ cm$^{-3}$ and b)
gas-dust heating around $10^5$ cm$^{-3}$.
Note that H$_2$O collisional de-excitation dominates the heating
for super-solar oxygen abundances. In fact, an EOS as stiff as in
Spaans \& Silk (2000) for dust of 100 K is obtained in this work.

The resulting IMF, as also argued
in Spaans \& Silk (2000) for solar abundances, would be top-heavy and
the overabundance of UV-luminous massive stars would continue the starburst
cycle through radiative feedback. It is interesting to note that the starburst
model (14) attains a value of $\gamma\approx 1.4$ at $n_{\rm H}\approx 10^4$
cm$^{-3}$. This is somewhat larger than the canonical value of
$\gamma_{\rm c}=4/3$ for {\it spherical} ($\nu =3$) clouds.

\section{Future Work}

There are a number of issues that should be addressed in the future:
I) Observations of water lines with the HIFI instrument on Herschel should be
performed to determine the H$_2$O level populations and substantiate our claim
of a polytropic index  $\gamma >1$ in starburst galaxies.
II) Numerical simulations with density and metallicity-dependent values of
$\gamma$, i.e., a pointwise polytropic EOS, should be performed
for collapsing gas clouds to assess the relative impacts of magnetic fields,
turbulence and the EOS on gravitational collapse and the shape
of the IMF (c.f.\ Klessen et al.\ 2004; Li et al.\ 2003). III) The difference
in influence of an
external FUV (O and B stars, popIII objects) or X-ray (QSOs, mini-quasars)
radiation field on the EOS of molecular clouds should be studied
(Meijerink \& Spaans 2004).

\acknowledgments

We would like thank Jan-Willem Pel for discussions on the nucleosynthesis
of massive stars and Dieter Poelman for help with the excitation of water.
JS gratefully acknowledges the hospitality of the Kapteyn Institute where this work was performed.

\end{document}